\documentclass[%
 reprint,
superscriptaddress,
 amsmath,amssymb,
prl,
]{revtex4-2}

\usepackage{graphicx}
\usepackage{dcolumn}
\usepackage{bm}

\begin{document}
\preprint{APS/123-QED}

\title{Integrate and scale: A source of spectrally separable photon pairs}
\author{Ben M. Burridge}
\email{benburridge@gmail.com}
\affiliation{Quantum Engineering Technology Laboratories, University of Bristol, Bristol, United Kingdom}
\affiliation{Quantum Engineering Centre for Doctoral Training, Centre for Nanoscience \& Quantum Information, University of Bristol, Bristol, United Kingdom}
\author{Imad I. Faruque}
\affiliation{Quantum Engineering Technology Laboratories, University of Bristol, Bristol, United Kingdom}
\author{John G. Rarity}
\affiliation{Quantum Engineering Technology Laboratories, University of Bristol, Bristol, United Kingdom}
\author{Jorge Barreto}
\affiliation{Quantum Engineering Technology Laboratories, University of Bristol, Bristol, United Kingdom}

\begin{abstract}
Integrated photonics is a powerful contender in the race for a fault-tolerant quantum computer, claiming to be a platform capable of scaling to the necessary number of qubits. This necessitates the use of high-quality quantum states, which we create here using an all-around high-performing photon source on an integrated photonics platform. We use a photonic molecule architecture and broadband directional couplers to protect against fabrication tolerances and ensure reliable operation. As a result, we simultaneously measure a spectral purity of $99.1 \pm 0.1$~\%, a pair generation rate of $4.4 \pm 0.1$~MHz mW$^{-2}$, and an intrinsic source heralding efficiency of $94.0 \pm 2.9$~\%. We also see a maximum coincidence-to-accidental ratio of $1644 \pm 263$. We claim over an order of magnitude improvement in the trivariate trade-off between source heralding efficiency, purity and brightness. Future implementations of the source could achieve in excess of $99~\%$ purity and heralding efficiency using state-of-the-art propagation losses.
\end{abstract}

\maketitle


\section{Introduction}
Progress toward practical quantum computational platforms is accelerating, with increasing competition from various industrial efforts \cite{Xanadu_Blueprint, PsiQ, Ion_FTQC, Google_QC}. Quantum computers promise the ability to efficiently simulate complex quantum systems \cite{Q_Chem, many_body}, solve classically infeasible cryptography challenges \cite{Shor}, and make possible revolutionary low-carbon technologies \cite{Batteries_Xanadu} amongst other applications. This is in direct contrast to the lengthy and comparatively inefficient computations of our classical methodologies \cite{Feynman_QC_CC}. Quantum photonics is one such platform, and further benefits from its mass-manufacturability (integrated photonics) on chip-scale devices \cite{Integrated_Quantum, Integrated_Quantum_2022}. From an applications perspective, it enables the likes of large-scale entangled quantum networks \cite{Entangled_Network8, Entanglement_MUX}, next-generation sensing \cite{Climate_Sensing}, and ultra-precise measurements \cite{Low_Noise}. The integrated photonics platform grows ever more promising with the developments of many fundamental building blocks needed for a photonic quantum computer \cite{Waveguide_Pure, low_loss_Si, Source_Mux, Cryo_Modulator, SNSPD_1350, SNSPD_1550}, and is possibly the only platform capable of reaching the number of qubits necessary for true fault-tolerance \cite{1MQubits}.

Much of the theory surrounding a photonic quantum processor postulates ideal quantum resources \cite{QC_Resource_Algorithm} for the purposes of their architecture \cite{FBQC, OWQC}, as well as for any error-correction schemes \cite{QC_Err_FT, QC_Err_Mem, QC_Err_Graph}. To create a photon source in line with these requirements we need high heralding efficiencies and high purities. High brightness sources are more of a practical requirement, as brightness determines the time frame on which a given computation can be performed \cite{Source_Mux}, with larger circuits needing brighter sources to maintain similar coincidence rates. Together, these metrics maximize the generation probability of sources; which is key for interfering photons from multiple sources, as well as maximizing any subsequent interference between them, making results like \cite{PQ_Adv} achievable. Recent examples targeting improved integrated source metrics showcase versatile \cite{BB_DP}, and bright sources \cite{Pure_Ring_MZI} of indistinguishable photons with limited heralding efficiency, or high-efficiency sources that trade brightness for purity, and are limited to broadband operation \cite{Waveguide_Pure} leaving them susceptible to additional noise inside the wider filtering band. 

Typically, the nonlinear process of spontaneous four-wave mixing (SFWM) is used to probabilistically generate pairs of photons (conventionally referred to as signals and idlers) on the CMOS-compatible silicon-on-insulator (SOI) platform. Devices can be designed to optimize the brightness of this process through long interaction lengths; manifesting as long sections of waveguide \cite{16D}, or high-field strengths in optical cavities \cite{Tele} which can more easily result in lower escape efficiencies \cite{Heralding_Rings}. These sources use the process of heralding \cite{OG_Herald} to mitigate the probabilistic nature of SFWM, with valid detection events occurring only with two or more coincident photons. The consequence of heralding is that, due to the energy and momentum-conserving nature of SFWM, any correlation between signal and idler photons projects the heralded photon into a mixed state. The heralded generation of entangled states \cite{Heralding} relies on the interference of multiple indistinguishable photons, and if the heralded single-photons are in a mixed state, this will degrade any quantum interference between them \cite{SPM_Purity}.

Here we showcase a photon-pair source using the resonantly enhanced SFWM process to maximize brightness and spectral purity, additionally, design optimizations of a photonic molecule architecture promise heralding efficiencies much higher than previously demonstrated for ring resonators. Our source comes with built-in resilience to fabrication variations to help ensure that these devices can be fabricated identically with high yields. The inevitable target will be to use large banks of these sources in parallel. This resilience comes in part from the strong-coupling regime that we operate in, but equally from our directional coupler designs. Motivated by previous works \cite{BB_DC_OG, BB_DC}, and using the transfer matrix method (TMM), we chose device geometries expected to offer broadband operation, and fabrication resilience, something that has been proposed \cite{BB_DC_OG} but never implemented in a full device to the knowledge of the authors.

\begin{figure}[t]
\centering
\includegraphics[width = \linewidth]{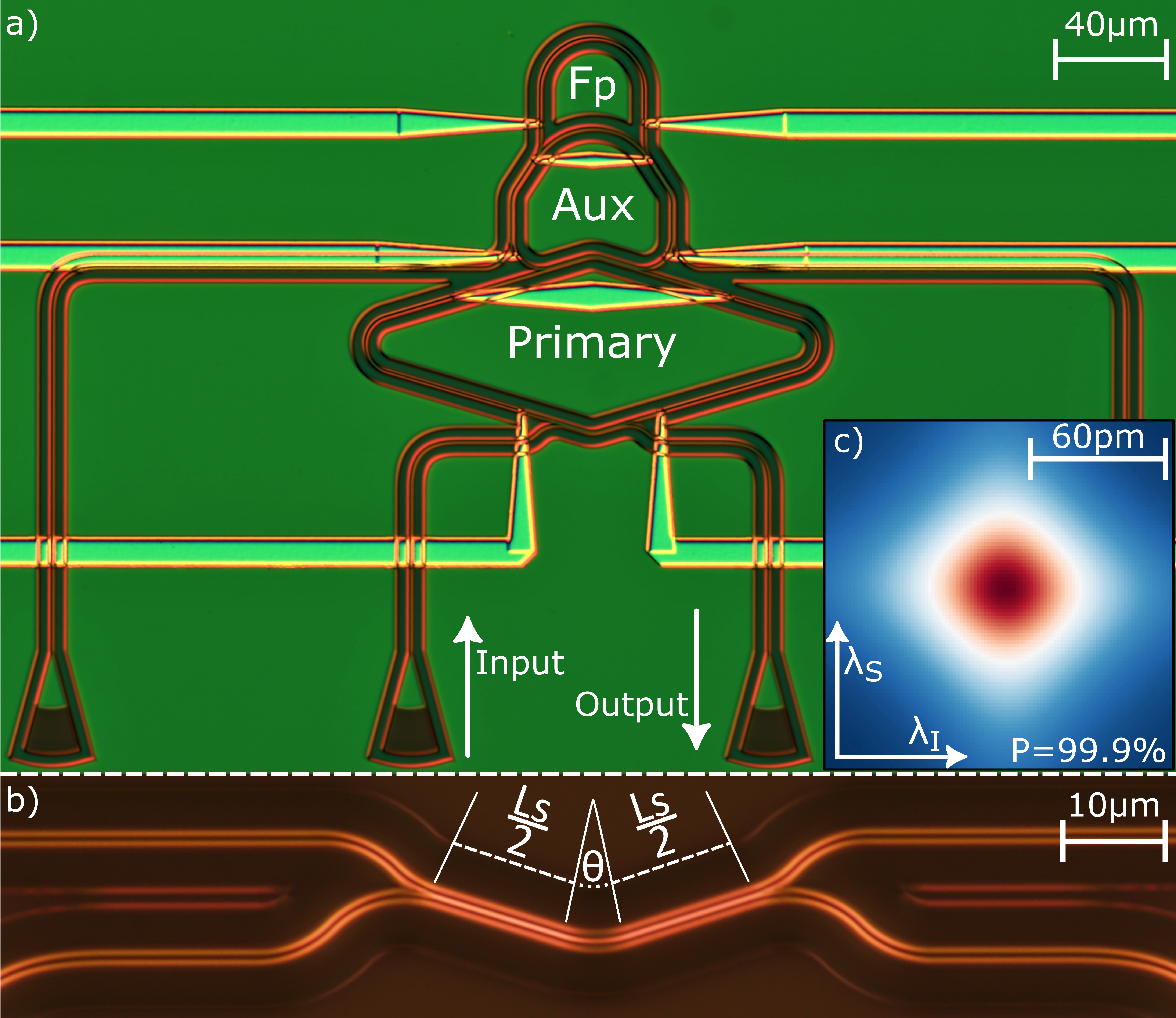}
\caption{Microscope images of: \textbf{a)} The photonic molecule photon-pair source. \textbf{b)} The bent directional coupler composed of straight coupled waveguides of length $L_S$, and bent coupled waveguides of length $L_C$ = $r\theta$, where $r$ is the coupler bend-radius, and $\theta$ is the angle swept by the bent region of the coupler. \textbf{c)} Simulated joint spectral intensity of our designed photonic molecule, where P is corresponding spectral purity, $\lambda_{S}$ and $\lambda_{S}$ are signal and idler wavelengths, respectively.}
\label{fig:Photonic_Molecule_Coupler}
\end{figure}

\section{Design}
Simple ring resonators are bound to purities of up to 91.7\% due to strict energy conservation conditions between identical linewidth resonances \cite{SFWM_Rings, AMZI_Ring_Theory}. However, this limit can be alleviated if the pump resonance linewidth is broadened relative to the signal/idler fields as in \cite{AMZI_Ring_Theory, Pure_Ring_MZI}, or alternatively, if the temporal response of the pump is sharpened \cite{DP_Theory, BB_DP}. Reducing the interaction length or amplitude of the pump field in any way will lead to a predictable decay in the brightness of the SFWM process \cite{AMZI_Ring_Theory, DP_Theory}.

Figure.~\ref{fig:Photonic_Molecule_Coupler}a shows our resonator-based photon source design that allows us to engineer the spectral response of the in-resonator pump fields compared to those of the signal/idlers. The coupled resonators (primary-auxiliary) mean that we engineer only the shared resonances between the rings, and restrict photon generation solely to the primary resonator. Promisingly, the idea of a photonic molecule has already proved a versatile and auspicious design \cite{SFWM_PM, PM_Dispersion, PM_Purity}. Here, our control over the in-resonator pump spectrum comes from the inter-resonator coupling between our primary resonator, which is simultaneously resonant at all three wavelengths of interest (signal, idler, and pump), and our auxiliary resonator which is solely resonant at our pump wavelength. Additionally, we implement a loss channel (\textbf{Fp}) for only the pump wavelengths using an asymmetric Mach-Zehnder interferometer (AMZI) to avoid inducing excess loss for the signal/idler photons. As part of the auxiliary resonator, \textbf{Fp} provides control over the auxiliary resonance linewidth and hence, control over the lineshape of the pump in the primary resonator, and for the purposes of source brightness, its field enhancement. In expanding the possible resonant lineshapes of our pump beyond that of a simple Lorentzian \cite{Backscatter}, we aim to reduce the trivariate trade-off between purity, brightness, and heralding efficiency that is prominently discussed in previous works \cite{Heralding_Rings, AMZI_Ring_Theory, DP_Theory}.

\begin{figure}[t!]
\centering
\includegraphics[width = \linewidth]{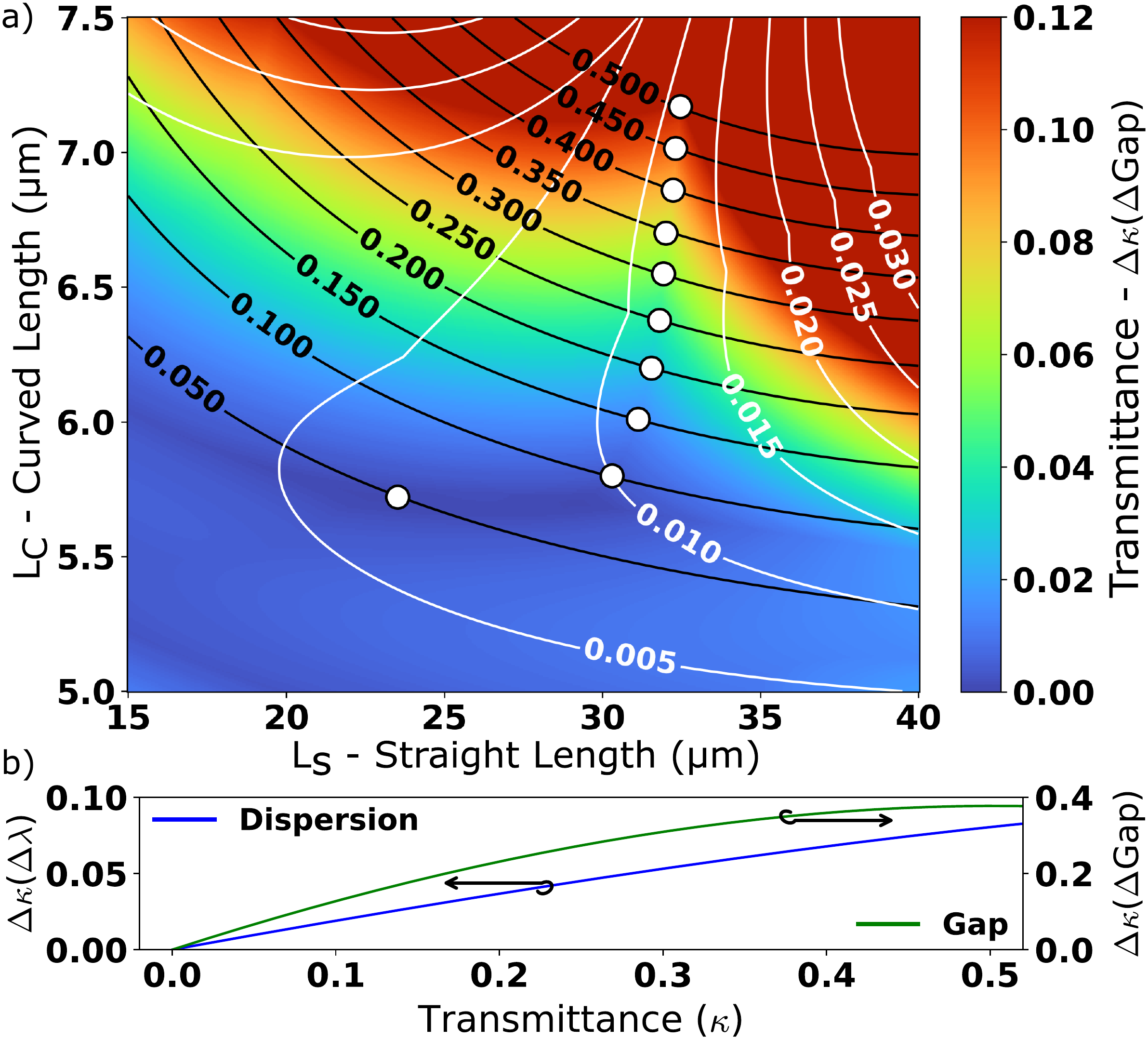}
\caption{\textbf{a)} Directional coupler design space that describes the change in transmittance of specific geometries of couplers for a waveguide separation of 200~nm~$\pm~\Delta$Gap (50~nm) and $\lambda$ = 1550~nm, as a measure of fabrication tolerance. The solid black contours denote specific transmittances in multiples of 0.05, and the solid white contours denote the maximum wavelength variance (dispersion) of the transmittance over the telecom c-band ($\Delta \lambda$ = 1530~nm - 1565~nm) compared to $\lambda$ = 1550~nm. Marked white points indicate the most fabrication-tolerant couplers at each transmittance. \textbf{b)} Fabrication tolerance and dispersion of a straight evanescent coupler against transmittance, for a waveguide separation of 300~nm and $\lambda$ = 1550~nm.}
\label{fig:DC_Metrics}
\end{figure}

\begin{figure*}[ht!]
\centering
\includegraphics[width = \linewidth]{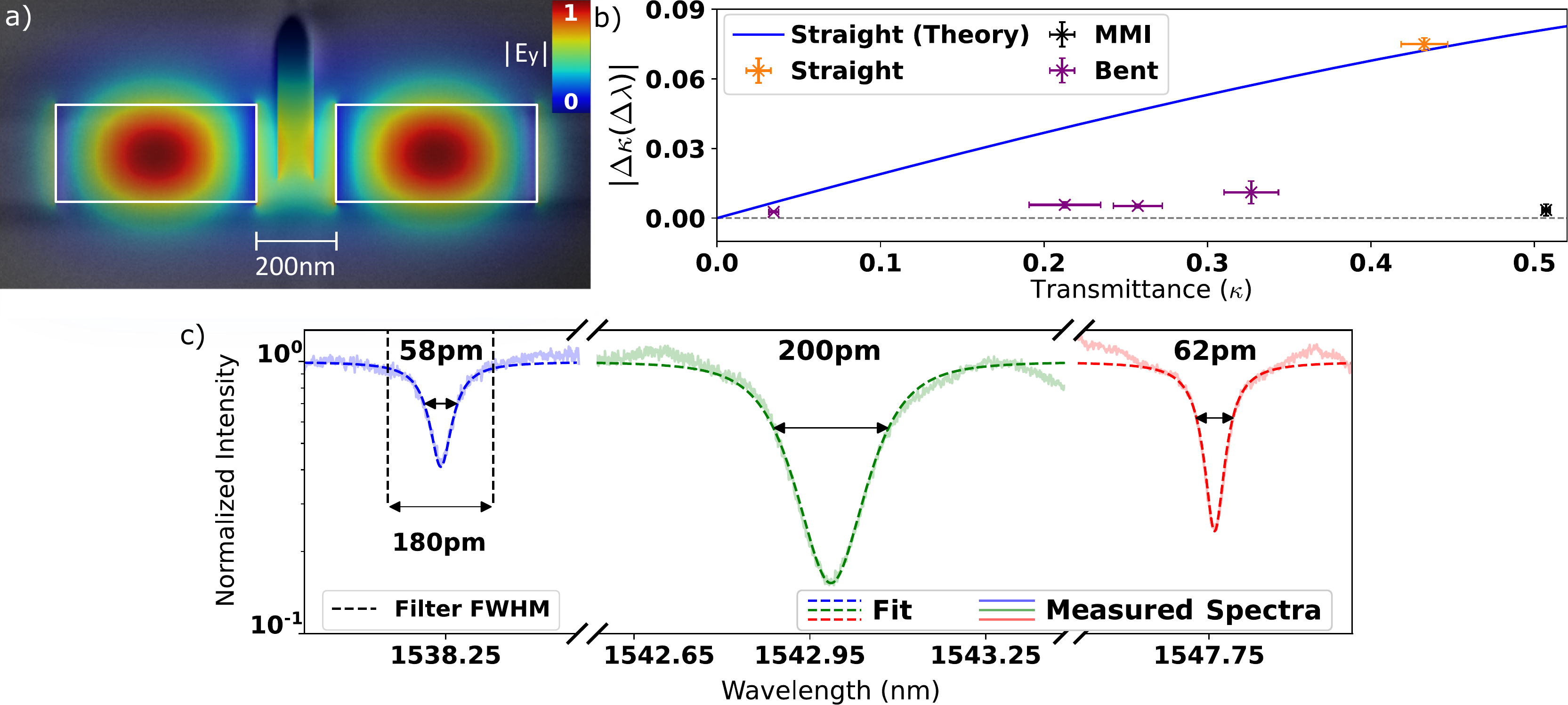}
\caption{\textbf{a)} Scanning electron microscope (SEM) images of a cross-section of our (bent) coupler. The cross-section was achieved by using a focused ion beam (FIB) to mill away the surrounding area to give access to the SEM. We overlay a simulation of the modal solution (see Supplement 1) inside a coupler of this approximate geometry (waveguides cores highlighted) to show the strong interaction with the cladding void. Waveguide separation is 200~nm, height is 220~nm, and width is 500~nm. \textbf{b)} Transmittance ($\kappa$) measurements of our bent coupler test structures at $\lambda$ = 1550~nm, and their maximum variance $\Delta \kappa ( \Delta \lambda)$ over the telecom c-band. Data for standard MMI and directional couplers are included, both targeting 3dB operation. \textbf{c)} Measured and fitted spectra of our photonic molecule's signal, pump, and idler resonances, respectively.}
\label{fig:Spectra_Couplers}
\end{figure*}

\begin{figure*}[ht]
\centering
\includegraphics[scale = 0.48]{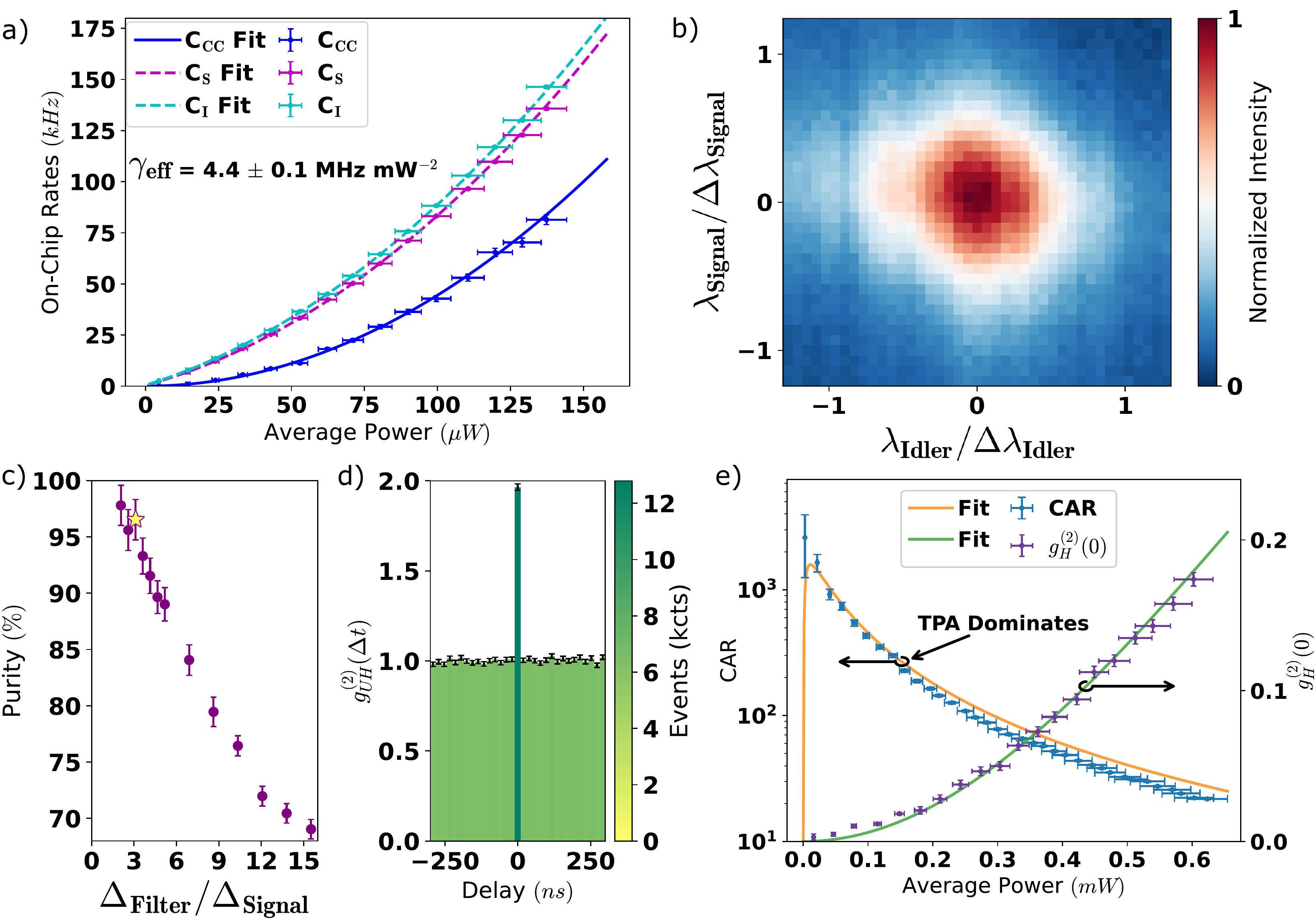}
\caption{\textbf{a)} Measured brightness of our photonic molecule. Fits of the raw data are used to obtain the brightness ($\gamma_{eff}$), and the heralding efficiencies ($\eta_{s, i}$) for the signal and idlers, respectively. On-chip rates have been inferred from our estimates of $\eta_{s, i}$. \textbf{b)} Measured joint spectral intensity of the source, $\Delta_{Idler, Signal}$ are the respective linewidths of the idler and signal resonances. \textbf{c)} Unheralded second-order correlation measurements for a range of filter bandwidths $\Delta_{Filter}$. \textbf{d)} Histogram for the second-order correlation function measurement for the starred data point in Figure.~\ref{fig:Source_Metrics}c. \textbf{e)} Measurements of the coincidence-to-accidental ratio (CAR) and heralded second-order correlation function, the point at which two-photon absorption (TPA) becomes non-negligible is annotated.}
\label{fig:Source_Metrics}
\end{figure*} 

The platform chosen for this work uses waveguide geometries of 500~x~220~nm, and a bend-radius of 10~$\mu$m as a compromise between footprint and bend-losses, with estimated propagation losses of $<$5~dB/cm. This allows our source to fit inside a footprint of 143~$\mu$m x 172~$\mu$m. 
In assessing the sensitivity of straight-directional couplers ($L_C = 0$) to both wavelength and fabrication variations (specifically waveguide separations of $\pm$~50~nm), we saw that higher transmittance lead to increased sensitivity in both categories (Fig.~\ref{fig:DC_Metrics}b). This is in stark contrast to the design space of the bent couplers (Fig.~\ref{fig:DC_Metrics}a) where both wavelength and fabrication sensitivity can be minimized by choosing a specific geometry of coupler. For our bent couplers, we chose a waveguide separation of 200~nm to minimize their footprint, as well as their wavelength and fabrication sensitivity. 

\section{Results and Discussion}
\subsection{Robustness}
Any practical photon-pair source must retain its mass-fabricability, and as such, it needs to be resilient to fabrication inaccuracies. The intrinsic design of our source plays a major role in its robustness as targeting higher heralding efficiency using a ring-resonator necessitates operating in the strongly over-coupled regime \cite{Heralding_Rings}, which, if targeting an all-around high-performance photon source, is one of many constraints we need to operate within. This regime comes with inherent tolerance to fabrication as small deviations will proportionally impact the design less. In particular, without \textbf{Fp} we would have to work with very small, and therefore sensitive couplings between the primary and auxiliary resonator. This is analogous to the limited purity gains (97\%) from a small amount of pump backscattering in ring resonators \cite{Backscatter}, in our design \textbf{Fp} allows us to work with much higher inter-ring coupling strengths, and also leads to higher purities (see Supplement 1).

As for the couplers themselves, SEM images of the devices (Fig.~\ref{fig:Spectra_Couplers}a) highlighted the limitations of the fabrication technology and confirmed the anticipated presence of voids (absence of cladding). Voids are a common issue when depositing thick films using chemical vapor deposition (CVD) on structures with aspect ratios larger than 1:1, and the likelihood of voids is increased inside devices with small waveguide separations \cite{Voids}. Accordingly, this can be alleviated by reducing the aspect ratio of the coupling region (ratio of waveguide height and separation --- \cite{Void_AR}), which is a means of adding some fabrication tolerance to straight directional couplers at the cost of device footprint. Regardless, we can infer their robustness by characterizing the coupler test structures across the chip that are located on average 2.9~mm apart to ensure no local correlations (Fig.~\ref{fig:Spectra_Couplers}b). Our devices for comparison were a straight directional coupler and a standard multimode interferometer (MMI) both for 3~dB splitting at 1550~nm. We measured each device's transmission across the telecom c-band to test their performance and saw almost negligible dispersion compared to a straight coupler (Fig.~\ref{fig:Spectra_Couplers}b). Each coupler performed well with a small spread in transmittance across the chip which fell within expectations (Fig.~\ref{fig:DC_Metrics}a) and the dispersion of each coupler continued to compete with that of an MMI (Fig.~\ref{fig:Spectra_Couplers}b). This is further evident in the spectrum of the photonic molecule (Fig.~\ref{fig:Spectra_Couplers}c --- full spectra in Supplement 1), where over $\sim$10~nm we see very little change in the linewidth of the resonances.  Interestingly, the straight coupler's transmittance is lower than expected for the device, which could still be explained by a cladding void that forms later in the PECVD process.

\subsection{Brightness}
We experimentally characterized our photon-pair source by pumping it with a 340~pm (9~ps) bandwidth pulsed laser at a 51~MHz repetition rate, to excite the pump resonance of the ring which has a bandwidth of 200~pm (Fig.~\ref{fig:Spectra_Couplers}c). To estimate source brightness we varied the power of the pulsed laser on-chip using a variable optical attenuator (VOA) and measured the dependence of coincidences and singles with on-chip power. On-chip average power was estimated using a 90:10 fiber coupler, and the insertion loss of our grating couplers (3.8dB --- see Supplement 1). We kept the pump power low to avoid excess nonlinear loss from two-photon absorption (TPA, \cite{TPA}) which is annotated in Figure.~\ref{fig:Source_Metrics}e where the expected coincidence-to-accidental ratio (CAR) deviates from the ideal fit at about 0.15~mW. We can solve for heralding efficiencies at the detectors and the effective nonlinearity ($\gamma_{eff}$) by quadratically fitting the singles and coincidence rates (Fig.~\ref{fig:Source_Metrics}a --- Supplement 1). The $\gamma_{eff}$ that we extract is 4.4~$\pm$~0.1~MHz~mW$^{-2}$. The effective nonlinearity essentially characterizes the on-chip generation rate, and therefore the brightness of the photon-pair source. By performing a similar measurement after taking the auxiliary ring off-resonance, we see that $\gamma_{eff}$ jumps to 15.5~$\pm$~0.4~MHz~mW$^{-2}$ (see Supplement 1). This 3.5x increase in brightness translates to a mere 1.9x increase in peak power delivered to the photonic molecule to achieve the same rates. Comparatively, \cite{AMZI_Ring_Theory} predicts that for our resonant setup, we would see a 46x decrease in brightness, corresponding to a 6.8x increase in peak power to recover the same generation rate of a single ring, demonstrating the power of our design with over an order of magnitude improvement in the expected brightness.







\subsection{Purity}
We characterized the source spectral purity using stimulated emission tomography (SET --- \cite{SET}) to extract the joint spectral intensity (JSI) of the photon-pair source, followed by a Schmidt decomposition on the corresponding joint spectral amplitude (JSA) to obtain an estimate of spectral purity. Additionally, we took measurements of the unheralded second-order correlation function ($g^{(2)}_{uh}$ --- \cite{unheralded_g2}) to corroborate our estimate of purity. The heralded $g^{(2)}_h$ is an additional metric that quantifies the photon-number purity and is due to the probabilistic nature of SFWM combined with our non-photon-number resolving detection. It is related to the likelihood of producing more than one photon pair in a single coincidence window. For our SET measurements, we used a continuous-wave (CW --- T100S-HP) laser with a linewidth of 3.2~$am$ to perform a wavelength scan (1~$pm$ resolution) over the signal resonance, measuring the output power at the idler resonance using a high-resolution (0.16~$pm$) optical spectrum analyzer (OSA --- Waveanalyzer 1500s). The result of this measurement is presented in Figure.~\ref{fig:Source_Metrics}b, and the corresponding purity is 99.1~$\pm$~0.1$\%$, in-line with the predictions of \cite{AMZI_Ring_Theory} and unequivocally demonstrating the unentangled nature of our photons. While the JSI discards phase information of the JSA, previous works \cite{BB_DP} have reported that for high-purity JSIs, the purity estimate of the JSI converges with that of the true JSA, for which our simulations suggest a difference of only $0.01\%$ (see Supplement 1). To obtain the error on our JSI measurement, we used a supersampling technique to take advantage of the huge precision afforded to us by our equipment, allowing us to reduce the effect of noise from both our OSA and our tunable CW laser. This reduced the uncertainty in our measurement, and at a final resolution of 4~$pm$, plateaued in both error and purity (see Supplement 1), allowing us to say with confidence that our measurement contains the true value of purity.

While $g^{(2)}_{uh}$ measurements can be subject to more noise due to its unheralded nature \cite{imad_g2}, it can serve to verify a source's spectral purity in a scenario more closely resembling a realistic use-case. We pass our pump through a 400~$pm$ bandwidth filter to ensure no excess noise contaminates our measurement. Taking our signal photons, we filter them to varying degrees using the tunable filter, send them into a 50:50 fiber beamsplitter, and measure the CAR between the two arms of the beamsplitter. This ratio should be 2 if the photon is in a pure state, i.e. exhibiting perfect thermal statistics. Our results are shown in Figure.~\ref{fig:Source_Metrics}c. The best measured $g^{(2)}_{uh}$ is 1.98~$\pm$~0.02 for a filter bandwidth twice as wide as our source, which ensures the highest removal of any broadband nonlinear noise surrounding the source due to the 100~$\mu m$ of waveguide either side of our source. We chose this length of input waveguide to provide a practical estimate of how this source could perform as part of a larger circuit according to our other works \cite{Hidden_Noise}. However, finding the best compromise between noise removal and filter bandwidth ($\Delta_{Filter}/\Delta_{Signal} = 3$) we observe a $g^{(2)}_{uh}$ of 1.97~$\pm$~0.02 (Fig.~\ref{fig:Source_Metrics}d) where we measure our highest CAR between signal and idler photons of 1644~$\pm$~263 (Fig.~\ref{fig:Source_Metrics}e). Only filtering to this extent also avoids any serious degradation of the heralding efficiency which becomes an issue for filters approaching the bandwidth of the source \cite{HE_Limits}. This measurement agrees with that of our measured JSI and further certifies the purity of our photons.
Finally, we measure a minimum $g^{(2)}_h$ value of 0.0029~$\pm$~0.0021 (Fig.~\ref{fig:Source_Metrics}e), well inside the single photon regime. 

\subsection{Efficiency}
To get the best idea of the source's full efficiency we use a filtering bandwidth of 500~pm. The heralding efficiency that we measure off-chip is $\eta_s$ = 7.2~$\pm$~0.2~$\%$, and $\eta_i$ = 5.6~$\pm$~0.2~$\%$, where $\eta_s,i$ indicate the efficiencies of the signal and idler photons, respectively. However, we care more about the intrinsic heralding efficiency of the source, because that is the limiting factor for any on-chip implementation of single-photon detectors, which would remove coupling-related insertion losses. We tally up the losses of our setup by measuring the insertion loss from the output of our laser power reference to our detectors both with and without the chip to isolate the insertion loss of the signal and idler channels. Most of our losses come from our filtering (See Supplement 1), which adds insertion losses of approximately 6~dB in total and consists of a fiber DWDM (1~$nm$ bandwidth) for coarse filtering, and more noise-isolating filtering using a tunable filter (XTA-50). Additionally, we have characterization data from the detectors we are using to be able to account for non-unity detection efficiencies equivalent to insertion losses of -1.060~$\pm$~0.044~dB, and -0.814dB~$\pm$~0.026~dB (See Supplement 1). After this analysis, the post-source heralding efficiencies that we estimate are $\eta_s$ = 92.1~$\pm$~3.2~$\%$, and $\eta_i$ = 94.0~$\pm$~2.9~$\%$, which is extremely competitive even with waveguide implementations of next-generation sources ($\eta$ = 91~$\pm$~9~$\%$ --- \cite{Waveguide_Pure}).

\section{Conclusions}
The photonic molecule architecture of our photon-pair source brings clear gains in all of the key metrics that are required to build a fault-tolerant photonic quantum computer. Our measured purities of 99.1~$\pm$~0.1$\%$, certified using $g^{(2)}_{uh}$ measurements, are fundamental in the creation of quantum resources through entangling operations \cite{PsiQ}. Additionally, with a competitive maximum measured heralding efficiency of 94.0~$\pm$~2.9~$\%$, especially when compared to previously reported pure resonators ($\eta$ = 52.4~$\%$ --- \cite{Pure_Ring_MZI}), our source successfully operates within on-chip loss thresholds of a fusion-based architecture \cite{FBQC}.
Our source surpasses expectations set by previous works \cite{BB_DP, DP_Theory, Pure_Ring_MZI, AMZI_Ring_Theory}, and reduces the trade-off between brightness, heralding efficiency, and purity, with a measured $\gamma_{eff}$ = 4.4~$\pm$~0.1~MHz~mW$^{-2}$, beating expected brightness degradation by over an order of magnitude \cite{AMZI_Ring_Theory}. Finally, our source is reasonably resilient fabrication defects (Fig.~\ref{fig:Spectra_Couplers}a) and variances, both through the implementation of fabrication-tolerant directional couplers with dramatically reduced dispersion, and the intrinsic over-coupled design of our source. 
Therefore, our results present the brightest, most efficient ring resonator source of pure photons to date, that has scalability at the core of its design. State-of-the-art propagation losses \cite{low_loss_Si} could improve source heralding efficiencies to values in excess of 99$\%$. The full power of our design can therefore be realized through the maturity of the fabrication process as all of our source metrics can only improve with reduced propagation losses due to higher Q-factors \cite{Heralding_Rings}, leading to a scalable and truly optimizable source. 





\section{Acknowledgments}
B.M.B. would like to thank Massimo Borghi, Will McCutcheon, and Gary Sinclair for useful discussions. The authors would like to thank Andy Murray for their technical assistance, as well as Laurent Kling and Stefano Paesani for their work characterising the efficiency of the detectors. The authors would also like to thank the team at CORNERSTONE, including Callum Littlejohns, Ying Tran, Mehdi Banakar, Martin Ebert,
James Le Besque, Georgia Mourkioti, and Eleni Tsanidou for their technical assistance and SEM imaging of our devices. The chip used in this work was fabricated using the facilities available at CORNERSTONE.

\section{Funding}
B.M.B. acknowledges the support of the EPSRC training grant EP/LO15730/1. The authors acknowledge the support of the EPSRC Quantum Communications Hub (EP/T001011/1) and Quantum Photonic Integrated Circuits (QuPIC) (EP/N015126/1). The authors also include the use of a paid MPW service CORNERSTONE 2 (EP/T019697/1).

\section{Supplemental document}
See Supplement 1 for supporting content. 

\bibliography{References}


\clearpage
\onecolumngrid

\begin{center}
\textbf{\Large Integrate and scale: A source of spectrally separable photon pairs\\
----- Supplemental Material -----}
\end{center}




\renewcommand{\thefigure}{S\arabic{figure}}
\setcounter{figure}{0}   

\section{Modal Simulations of Directional Couplers}

\begin{figure}[h!]
	\centering
	\includegraphics[width=1.0\columnwidth]{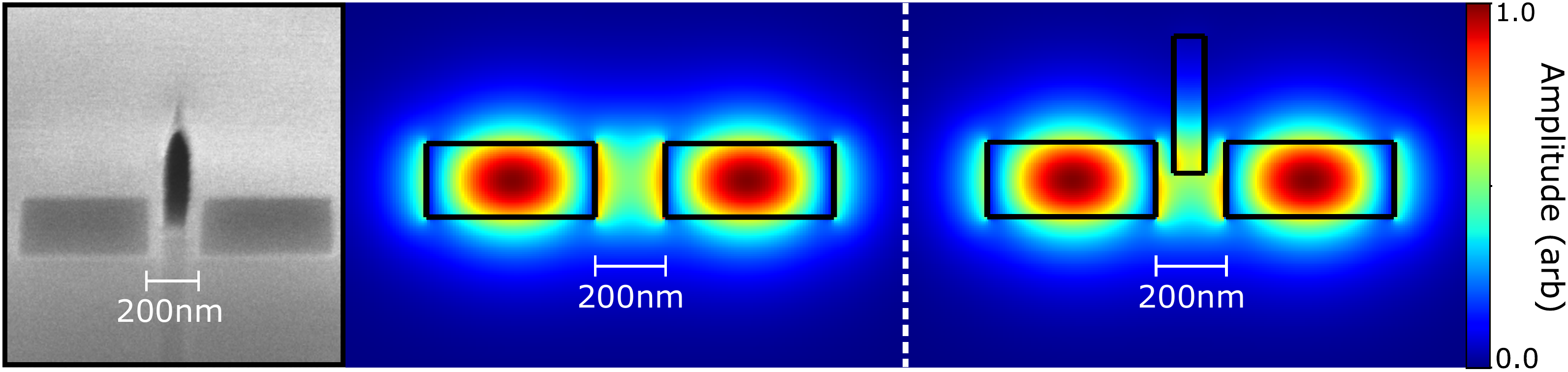}
	\caption{Simulations of the field ($E_y$) inside the straight section of a bent directional coupler without a cladding void (left) and with a cladding void (right).}
	\label{fig:Straight_Coupler_Voids}
\end{figure}

The methodology for the simulations of the bent directional coupler directly follows from the work in \cite{BB_DC}, and uses the transfer matrix method (TMM) to work out the final transmittance of a given coupler. To do this, we must solve the modes for each section of the directional coupler, meaning 6 solutions in total, 2 for each section (single waveguides, straight section, bent section). 
In doing this we construct the simulated waveguides exactly as we expect them to be fabricated.
The benefit of this is that very quickly we have access to the transmittance of any bent directional coupler within a range of straight and curved lengths ($L_s, L_c$). However, if the fabricated waveguides differ in some way, then of course the simulation will not accurately re-produce their real-world performance. This is where we have used SEM images of the devices provided by CORNERSTONE to give us as accurate a picture as possible for post-fabrication analysis. In these images, we observed the presence of cladding voids in the coupling region of our device (Fig.~\ref{fig:Straight_Coupler_Voids})

\begin{figure}[h!]
	\centering
	\includegraphics[width=0.6\columnwidth]{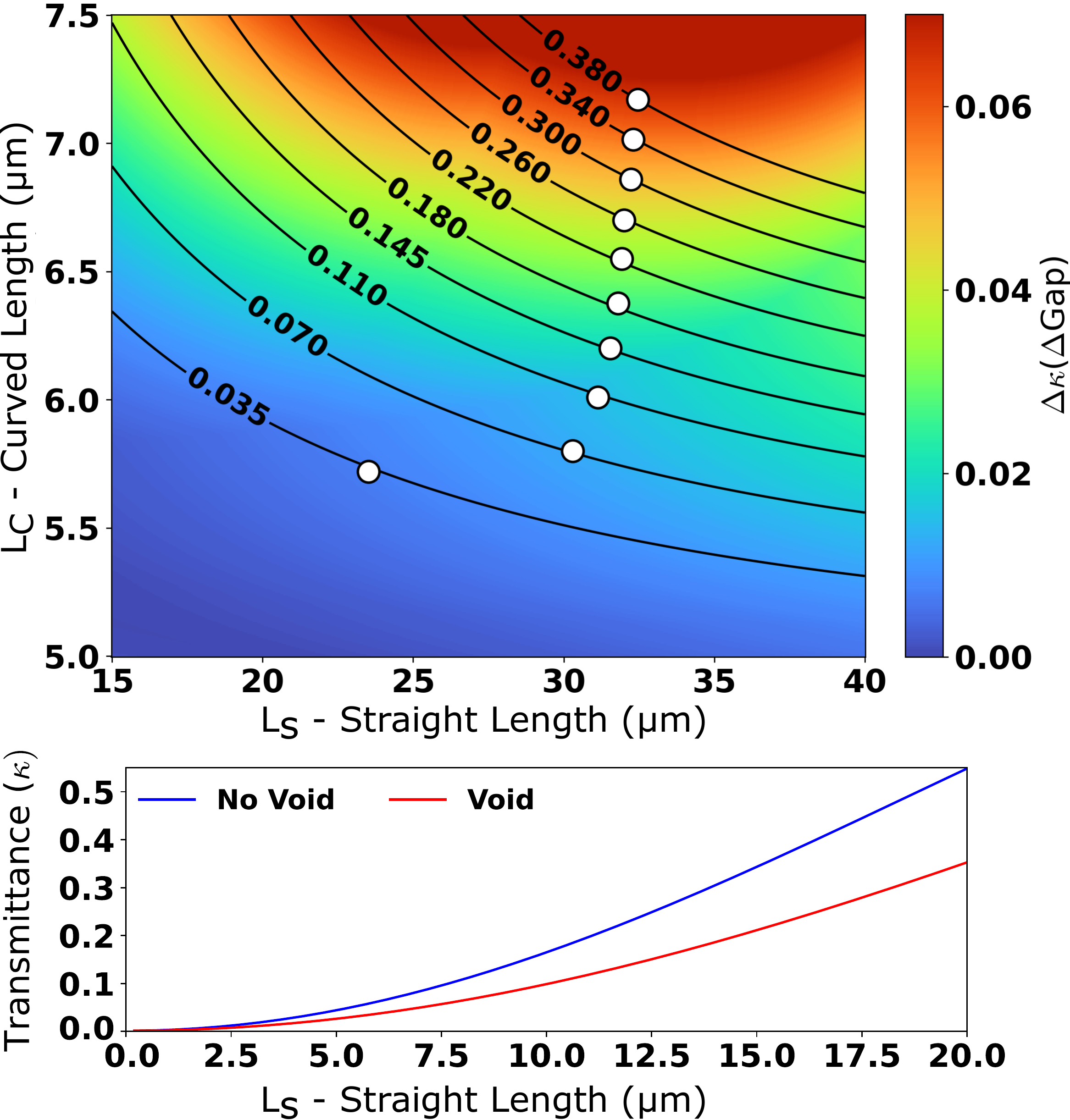}
	\caption{Upper) Corner-analysis based on coupler gaps of 185$\pm$20~nm, and void widths of 90$\pm$20~nm, with the heatmap indicating the maximum change in coupler transmittance ($\Delta \kappa (\Delta$Gap)). The black contours represent the transmittances for the mean gap and void dimensions that intersect with the original geometries of couplers (white marked points) plotted in the main text (Fig.~\ref{fig:DC_Metrics}). Lower) The dependence of transmittance of the length of a standard directional coupler, with and without a cladding void, for a waveguide separation of 200~nm.}
	\label{fig:BDC_DC_Comparison_Void_200nm}
\end{figure}

\begin{figure}[h!]
	\centering
	\includegraphics[width=\columnwidth]{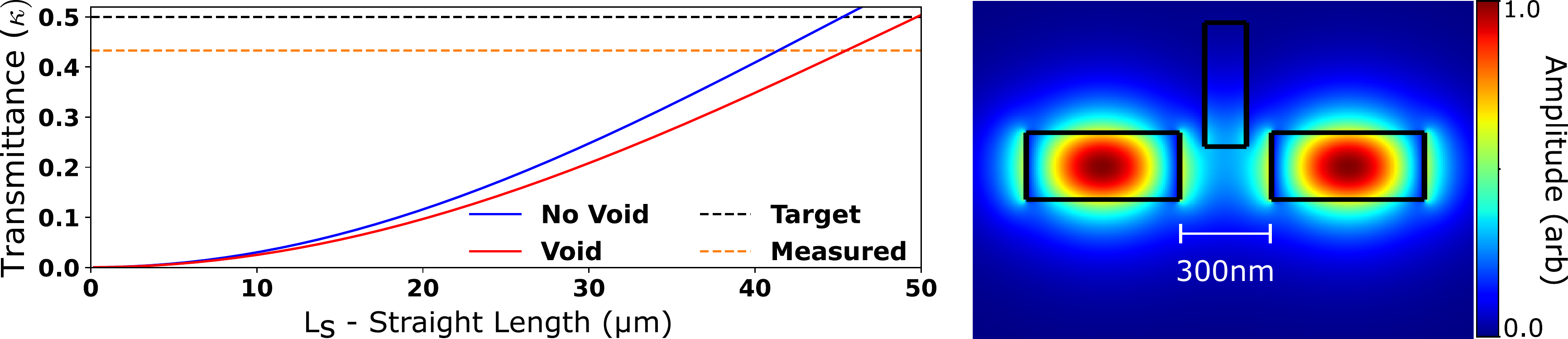}
	\caption{Simulations of a standard straight directional coupler, with waveguide separation of 300nm. Left) The dependence of transmittance on coupler length for couplers with and without cladding voids. Dashed lines represent the target and measured transmittance of our fabricated reference directional coupler. Right) Simulated mode profile ($E_y$) of the straight coupler with a cladding void.}
	\label{fig:Straight_Coupler_Voids_300nm}
\end{figure}

Solving for all of the modes using this new waveguide assembly, we saw that the performance of the bent couplers, while expected to be lower, was not so low as to be catastrophic. Especially when compared to straight directional couplers with a similar waveguide separation (Fig.~\ref{fig:BDC_DC_Comparison_Void_200nm}). When designing straight directional couplers for this chip, we used a waveguide separation of 300~nm for an added degree of fabrication tolerance. Looking at simulations of a cladding void in the centre of these couplers, they do seem to have a diminished effect (Fig.~\ref{fig:Straight_Coupler_Voids}). The behavior of Fig.~\ref{fig:Straight_Coupler_Voids_300nm} does a good job of explaining the underperformance of our straight coupler as mentioned in the main text. 

\clearpage

\section{Design Optimizations of a Photonic Molecule}

To effectively simulate our photonic molecule, first, we must solve our system of ring resonators to work out their transmission spectra and corresponding field enhancements. These field enhancements can be used to work out the performance of our source, as usual, \cite{SFWM_Rings}.

\begin{figure}[h!]
	\centering
	\includegraphics[width=0.68\columnwidth]{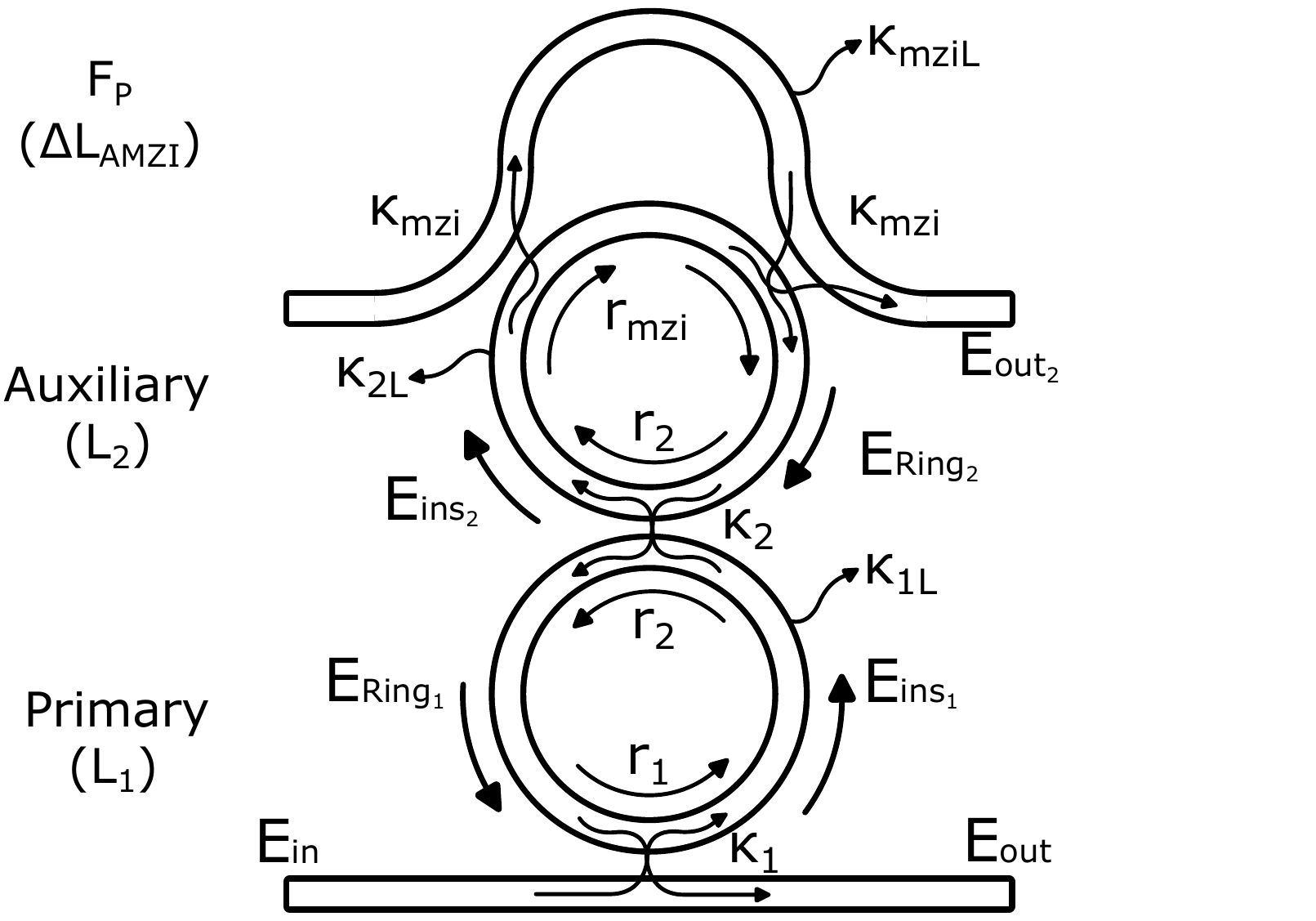}
	\caption[Photonic Molecule Schematic]{Schematic Drawing of the Photonic Molecule. The primary resonator has length $L_1$, the Aux resonator has length $L_2$ and the path difference inside $F_P$ is $\Delta L_{AMZI}$. The pump field ($E_{in}$) enters through the bus channel, before coupling to the Primary ring through $\kappa_1$ giving rise to $E_{ins1}$. The field inside the Primary will see $\kappa_2$, generating $E_{ins2}$ in the Auxiliary ring. $F_P$ acts as a wavelength-dependent loss channel for the auxiliary ring. Here the phantom loss channels are implied by the field transmission coefficients $\kappa_{1L}$, $\kappa_{2L}$ and $\kappa_{mziL}$.}
	\label{fig:Photonic_Molecule_Schematic}
\end{figure}

Our ring system can be schematically drawn as in Figure.~\ref{fig:Photonic_Molecule_Schematic}, and we can start our modelling similarly to the single-ring case. For the primary-bus coupling junction, we have

\begin{align}
\left(\begin{matrix}
E_{ins1}\\
E_{out}
\end{matrix}\right) 
= 
\left(\begin{matrix}
r_1 & i\kappa_1\\
i\kappa_1 & r_1
\end{matrix}\right)
\left(\begin{matrix}
E_{Ring1}\\
E_{in}
\end{matrix}\right)
\label{Eq:Dual_Ring_Bus_Scattering_Matrix}
\end{align}
where for simplicity $E_{Ring1}$ is the field before the coupling region $\kappa_1$.
We can write a similar set of equations for the primary-auxiliary coupling

\begin{align}
\left(\begin{matrix}
E_{ins2}\\
e^{\frac{-ikL_1}{2}}E_{Ring1}
\end{matrix}\right) 
= 
\left(\begin{matrix}
r_2 & i\kappa_2\\
i\kappa_2 & r_2
\end{matrix}\right)
\left(\begin{matrix}
E_{Ring2}\\
r_{1L}e^{\frac{ikL_1}{2}}E_{ins1}
\end{matrix}\right),
\label{Eq:Dual_Ring_Ring_Scattering_Matrix}
\end{align}
where $r_{1L}$ is the ring self-coupling due to the loss channel as before. Finally, we can write another set of equations for the $F_P$-auxiliary coupling 

\begin{align}
\left(\begin{matrix}
E_{out2}\\
e^{\frac{-ikL_2}{2}}E_{Ring2}
\end{matrix}\right) 
= 
\left(\begin{matrix}
r_{mzi} & i\kappa_{mzi}\\
i\kappa_{mzi} & r_{mzi}
\end{matrix}\right)
\left(\begin{matrix}
r_{mziL}e^{ik\Delta L_{AMZI}} & 0\\
0 & 1
\end{matrix}\right)
\left(\begin{matrix}
r_{mzi} & i\kappa_{mzi}\\
i\kappa_{mzi} & r_{mzi}
\end{matrix}\right)
\left(\begin{matrix}
0\\
r_{2L}e^{\frac{ikL_2}{2}}E_{ins2}
\end{matrix}\right),
\label{Eq:Dual_Ring_Fp_Scattering_Matrix}
\end{align}
where we have assumed no light enters the structure other than from $E_{in}$, and for simplicity have concatenated the relative phase of $F_P$ using the path-difference $\Delta L_{AMZI}$.
Now we can begin solving this system of equations, starting with Eq.~\ref{Eq:Dual_Ring_Fp_Scattering_Matrix}

\begin{align}
E_{Ring2} = \left(r_{mzi}^2 - \kappa_{mzi}^2r_{mziL}e^{ik\Delta L_{AMZI}}\right) r_{2L}e^{ikL_2}E_{ins2},
\label{Eq:E_Ring2}
\end{align}
followed by Eq.~\ref{Eq:Dual_Ring_Ring_Scattering_Matrix}
\begin{align}
E_{Ring1} = i\kappa_2e^{\frac{ikL_1}{2}}E_{Ring2} + r_2r_{1L}e^{ikL_1}E_{ins1},
\label{Eq:E_Ring1}
\end{align}
which we can use to start solving Eqs.~\ref{Eq:Dual_Ring_Bus_Scattering_Matrix} and \ref{Eq:Dual_Ring_Ring_Scattering_Matrix}

\begin{align}
E_{ins1} = \frac{i\kappa_2r_1e^{\frac{ikL_1}{2}}E_{Ring2} + i\kappa_1E_{in}}{ 1 - r_1r_2r_{1L}e^{ikL_1}},
\label{Eq:E_ins1_Start}
\end{align}
such that 
\begin{align}
E_{ins1} = \frac{i\kappa_2r_1e^{\frac{ikL_1}{2}}\left(r_{mzi}^2 - \kappa_{mzi}^2r_{mziL}e^{ik\Delta L_{AMZI}}\right) r_{2L}e^{ikL_2}E_{ins2}+ i\kappa_1E_{in}}{ 1 - r_1r_2r_{1L}e^{ikL_1}}.
\label{Eq:E_ins1_Finish}
\end{align}
At this point, I will define $r_2T_{Ring2} = r_2\frac{E_{Ring2}}{E_{ins2}}$ to represent the complex round-trip transmission of the auxiliary ring. Letting us re-write Eq.~\ref{Eq:E_ins1_Finish} as
\begin{align}
E_{ins1} = \frac{i\kappa_2r_1e^{\frac{ikL_1}{2}}T_{Ring2}E_{ins2}+ i\kappa_1E_{in}}{ 1 - r_1r_2r_{1L}e^{ikL_1}},
\label{Eq:E_ins1_Finish_Simp}
\end{align}
where the only unknown remaining is $E_{ins2}$ which we can now solve using Eq.~\ref{Eq:Dual_Ring_Ring_Scattering_Matrix}
\begin{align}
E_{ins2} \ = \ r_2T_{Ring2}E_{ins2} + i\kappa_2r_{1L}e^{\frac{ikL_1}{2}}\frac{i\kappa_2r_1e^{\frac{ikL_1}{2}}T_{Ring2}E_{ins2} + i\kappa_1E_{in}}{ 1 - r_1r_2r_{1L}e^{ikL_1}}, 
\label{Eq:E_ins2_Start}
\end{align}
which we can re-arrange to get
\begin{align}
E_{ins2} \ = \ \frac{-\kappa_1\kappa_2r_{1L}e^{\frac{ikL_1}{2}}E_{in}}{ \left(\left(1 - r_1r_2r_{1L}e^{ikL_1}\right) \left(1 - r_2T_{Ring2} \right) + \kappa_2^2r_{1L}r_1e^{ikL_1}T_{Ring2}\right) }.
\label{Eq:E_ins2_Finish}
\end{align}

Finally, using Eq.~\ref{Eq:Dual_Ring_Bus_Scattering_Matrix} we can work out the transmitted field of the photonic molecule as 
\begin{align}
E_{out} \ = \ i\kappa_1 E_{Ring1} + r_1E_{in},
\label{Eq:PM_Transmission_Start_1}
\end{align}
which is given by 
\begin{align}
E_{out} \ = \  i\kappa_1r_2r_{1L}e^{ikL_1}E_{ins1} - \kappa_1\kappa_2e^{\frac{ikL_1}{2}}T_{Ring2}E_{ins2} + r_1E_{in}.
\label{Eq:PM_Transmission_Start_2}
\end{align}

With a complete model, we can move on to simulating the performance of different designs of the photonic molecule. 

\begin{figure}[h!]
	\centering
	\includegraphics[width=0.9\columnwidth]{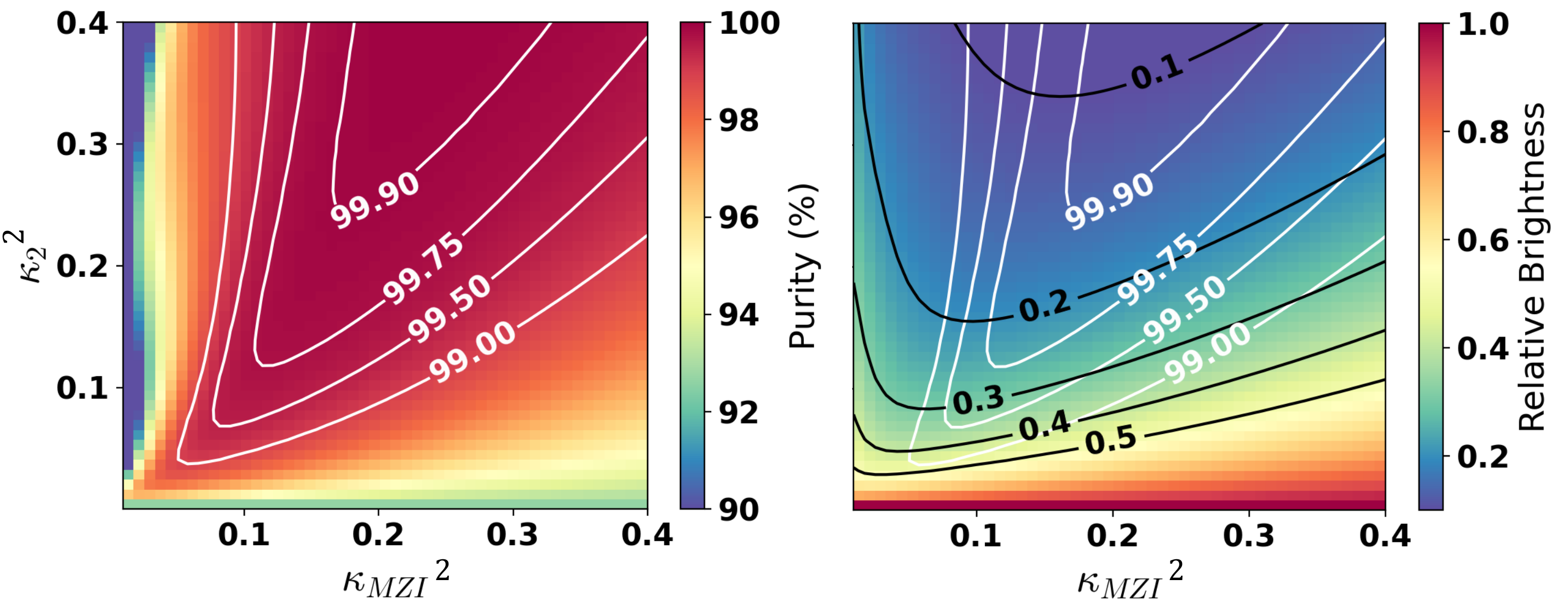}
	\caption{Scanning the primary-aux coupling strengths ($\kappa_{2}^2$), alongside the aux - mzi couplings ($\kappa_{MZI}^2$). Here we use $\alpha_{dB~cm^{-1}}$ = 3~$dB~cm^{-1}$, and $\kappa_{1}^2$ = 0.23. We model both the spectral purity (left), and the brightness (right). White contours indicate levels of constant purity, and black contours highlight levels of constant brightness.}
	\label{fig:Photonic_Molecule_Optimise_Aux_Fp}
\end{figure}

In designing our photonic molecule, the parameter we wanted to constrain ourselves with first was heralding efficiency. We can choose heralding efficiency using only one parameter, the ring-bus coupling of the primary resonator \cite{Heralding_Rings}. This necessitated working in the strong coupling regime, giving rise to our relatively broad 60~pm resonances of the primary resonator. By setting this constraint earlier, we only have a 2D parameter space to explore, which is the transmittance of the couplers used inside \textbf{Fp}, and the primary-auxiliary ring coupling.
Scanning these parameters allows us to quickly decide on an optimal photonic molecule for a given operating scenario (Fig.~\ref{fig:Photonic_Molecule_Optimise_Aux_Fp}).

We mentioned in the main text that the use of \textbf{Fp} allowed us to operate with higher coupling strengths, and this can be seen in Fig.~\ref{fig:Photonic_Molecule_Optimise_Aux_Fp}. Additionally, the features on these optimization plots are slowly varying at higher coupling strengths which highlight the fabrication tolerance of this particular source architecture. We can also make inferences about the brightness of our device. Given a single ring that operates at a brightness around an easily achievable 10~MHz, we maintain MHz rates well into the 99.9\% purity regime. Our source operates at a relative brightness of 0.28$\times$ that of a single ring, which is close to predicted.

\begin{figure}[h!]
	\centering
	\includegraphics[width=0.75\columnwidth]{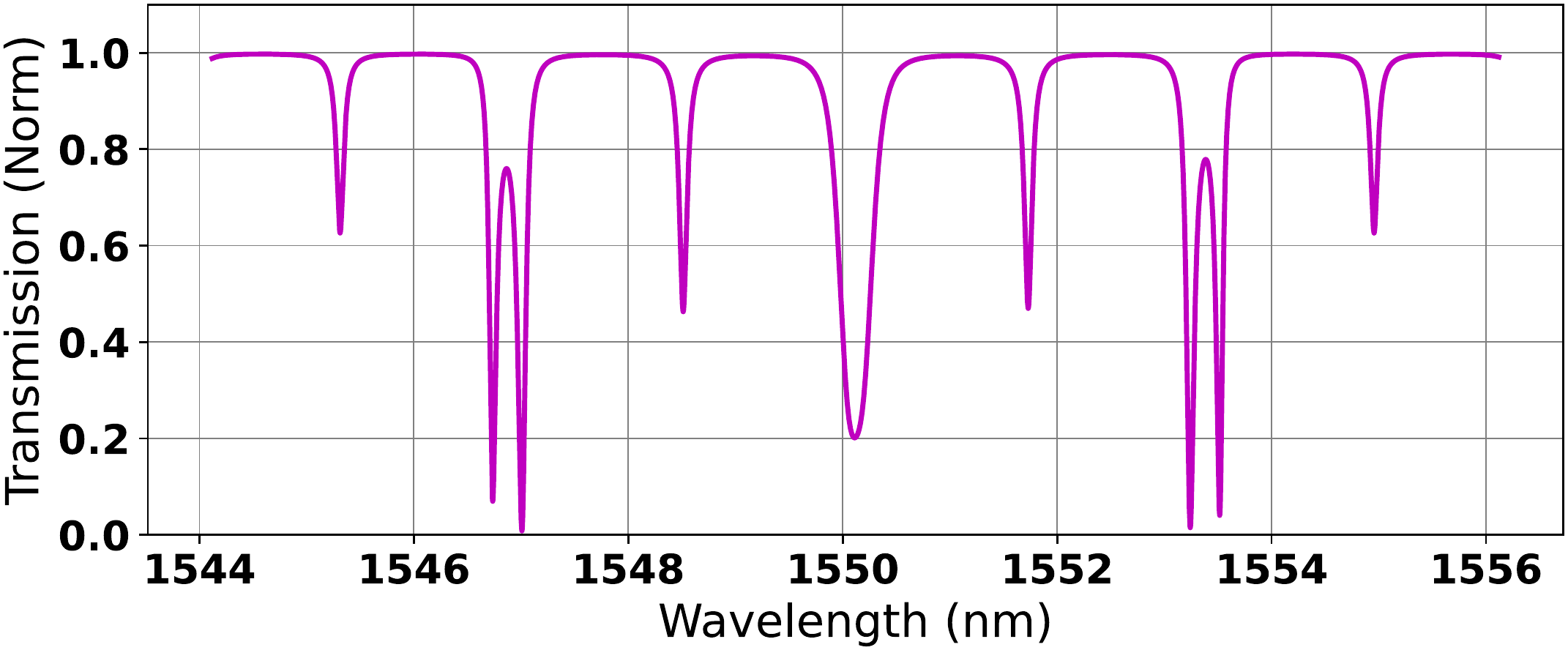}
	\caption{Simulated transmission spectrum of the photonic molecule that we simulated in the main text (Fig.~1c).}
	\label{fig:HH_Ring_Spectra_Theory}
\end{figure}

\begin{figure}[h!]
	\centering
	\includegraphics[width=0.75\columnwidth]{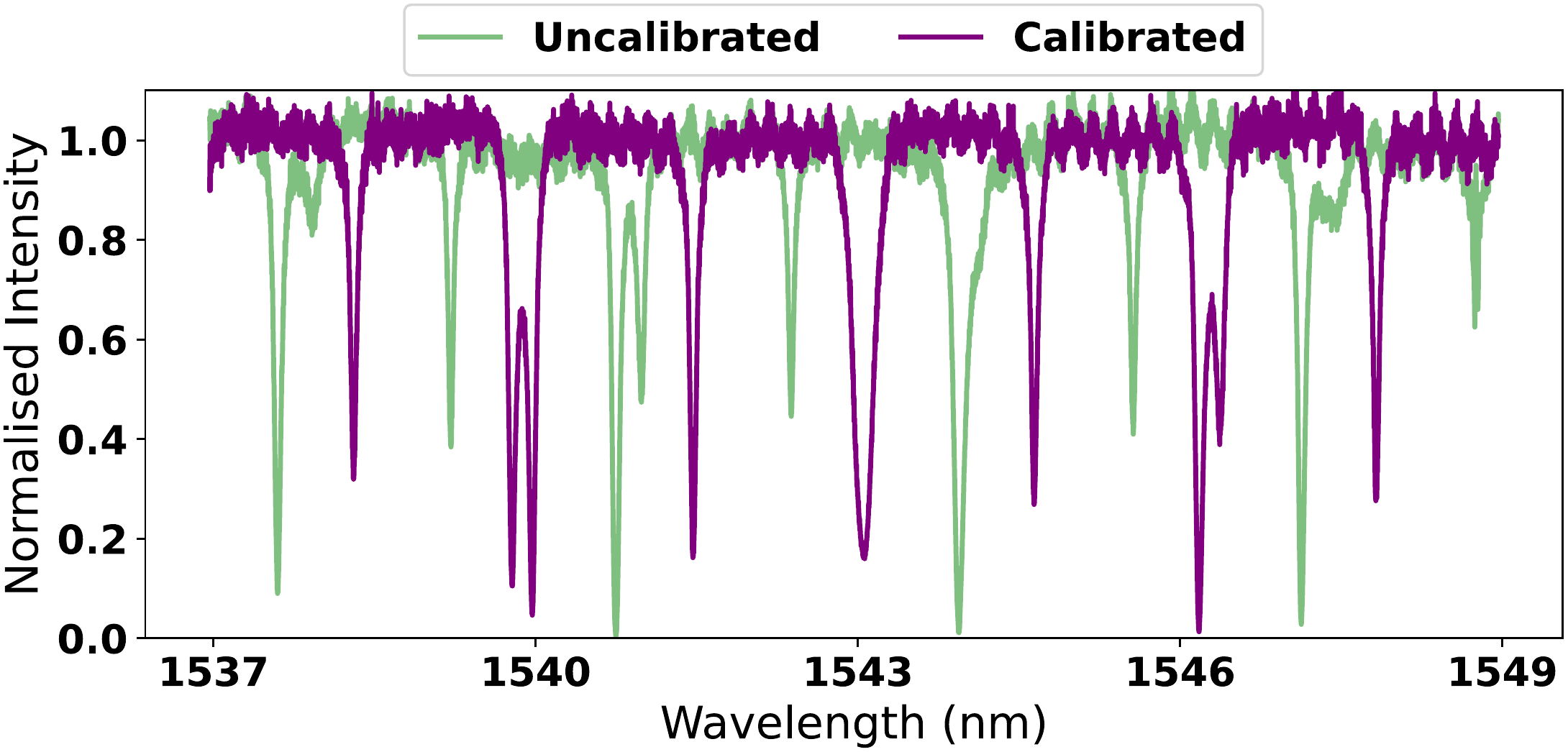}
	\caption{Experimentally recorded spectra of our photonic molecule, for both the uncalibrated and calibrated device. When uncalibrated, the photonic molecule does not have the same symmetric resonance broadening seen in the calibrated spectra.}
	\label{fig:HH_Ring_Spectra_Exp}
\end{figure}

\clearpage

\section{Brightness Measurements of a Photon-Pair Source}

To measure the brightness of our source, we correlate the rate of singles for each channel with the total true coincidence rate. We can do this based on our knowledge of the system, in that coincidence events ($C_{cc}$) occur between two photons of a pair through spontaneous four-wave mixing or random chance (accidentals -- $ACC$), single photon detections can come from SFWM events ($C_{s, i}$) and random linear noise which can be a result of pump leakage or other scattering events. Dark count rates ($DC$) of the detectors will also act as a constant offset to single photon detection rates. Using this knowledge, we fit the brightness curves using the equations

\begin{align}
C_{i} = \gamma_{eff}\eta_{i}P^2 + \beta_{i}P + DC, \label{Eq:Singles_1}\\
C_{s} = \gamma_{eff}\eta_{s}P^2 + \beta_{s}P + DC, \label{Eq:Singles_2}\\
C_{cc} = \gamma_{eff}\eta_{s}\eta_{i}P^2 + ACC, \label{Eq:Coincidences}\\
ACC = C_{i}C_{s} \Delta t, \label{Eq:Accidentals}
\end{align}
where $\gamma_{eff}$ is the brightness or effective nonlinearity of our pair source in Hz~mW$^{-2}$. The lumped linear noise coefficient for each channel is represented using $\beta_{i,s}$, and $P$ is the on-chip pump power.
These equations also allow us to extract the total heralding efficiency of the signal ($\eta_s$) and the idler ($\eta_i$) channels. In simpler terms, it is the on-chip pair generation rate (PGR), at 1~mW. The pair generation rate is our main point of comparison between sources, but PGR can be specified at any on-chip power, which then needs to be scaled appropriately for comparison. Finally, the accidentals ($ACC$) can be estimated from the detected singles rates as in Eq.~\ref{Eq:Accidentals} over the coincidence window ($\Delta t$) or, more typically, are calculated from the measured coincidence data. 

\begin{figure}[h!]
	\centering
	\includegraphics[width=0.5\columnwidth]{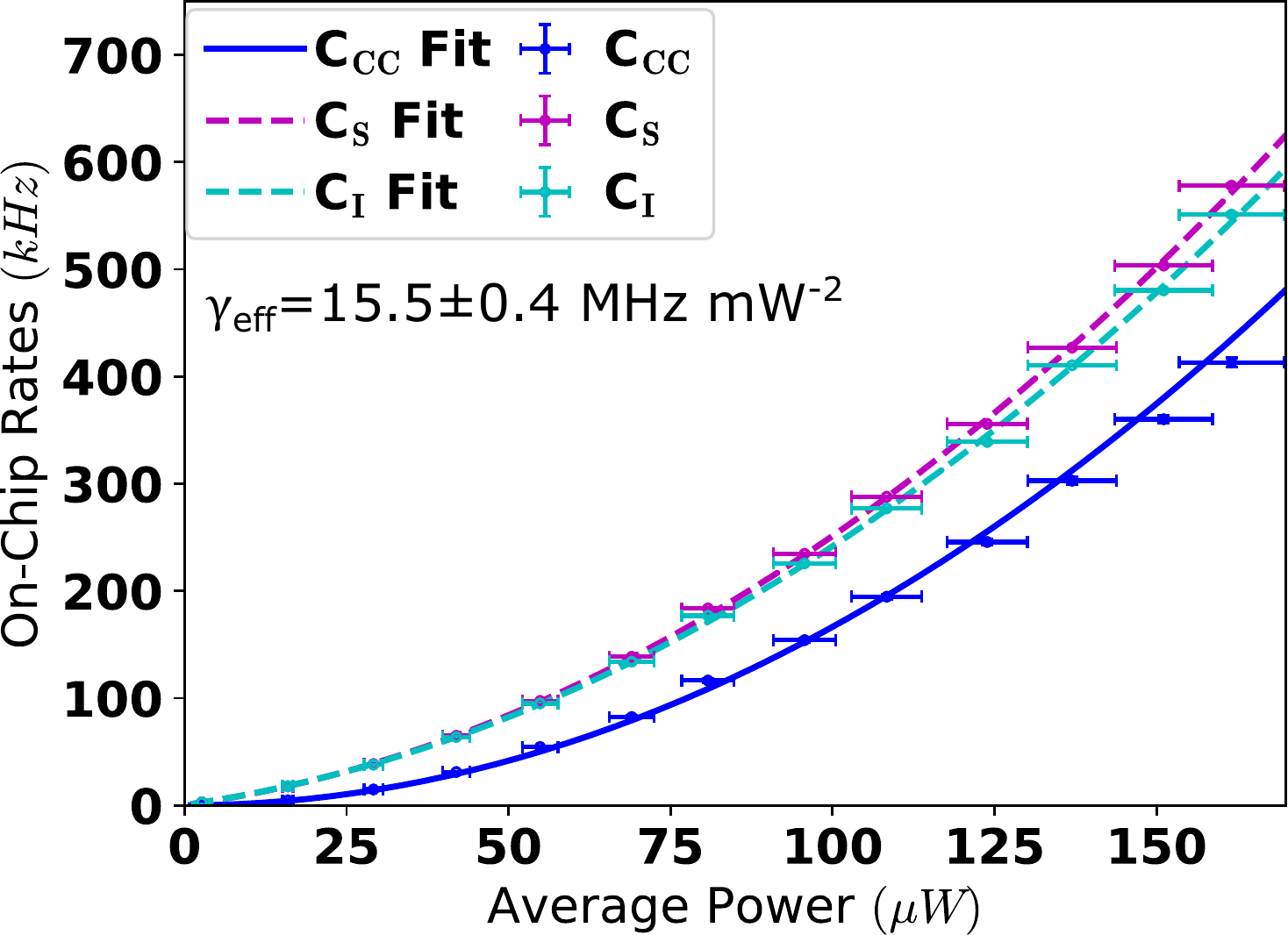}
	\caption{Brightness measurements when taking the auxiliary ring off-resonance.}
	\label{fig:Brightness_off_res}
\end{figure}

As mentioned in the main text, we also measured the brightness of the source if we took the auxiliary ring off-resonance, which gave us an estimate of the single-ring brightness of our particular setup. We measured the $\gamma_{eff}$ in this case to be 15.5~$\pm$~0.4~MHz mW$^{-2}$ (Fig.~\ref{fig:Brightness_off_res}) as mentioned in the main text. Fitting each of the curves according to Eqs.~\ref{Eq:Singles_1} - \ref{Eq:Accidentals} is what allows us to extract $\gamma_{eff}$.

\clearpage

\section{Correcting for Intrinsic Source Heralding Efficiency}

With the estimated heralding efficiencies from Eqs.~\ref{Eq:Singles_1} - \ref{Eq:Accidentals}, we want to work backwards to calculate the source's intrinsic heralding efficiency. To do this, we need to work out all of the losses from the source to the detectors, as well as the detection efficiencies for each channel. 

\begin{figure}[h!]
	\centering
	\includegraphics[width=0.99\columnwidth]{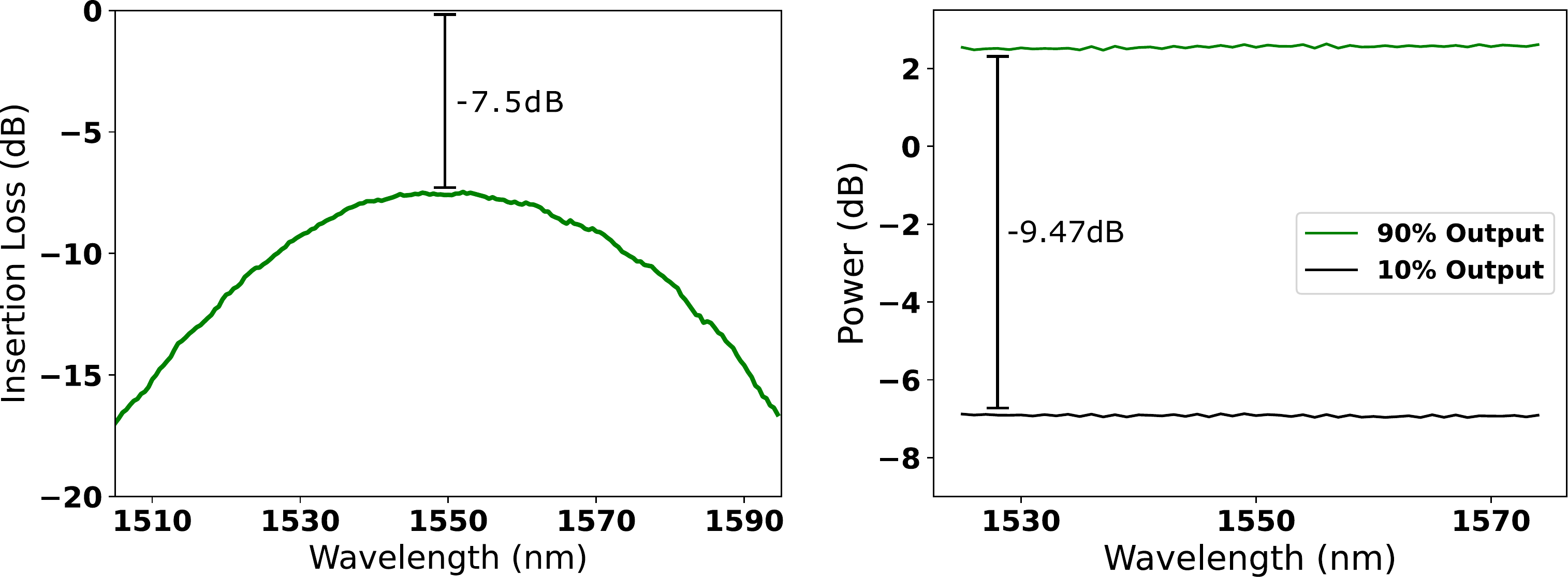}
	\caption{Left) Grating coupler spectrum from a small loop back structure, here we see a peak insertion loss of $-7.5~dB$, corresponding to $-3.75~dB$ per coupler. Right) The output power of our laser through the 90:10 fibre coupler that we use to measure the incident pump power to the chip.}
	\label{fig:Grating_Coupler_Spectrum}
\end{figure}

We can start by working out the loss due to our grating couplers, which we can estimate using a short loopback structure, and a 90:10 fibre coupler. The loopback structure consists of two grating couplers, and enough waveguide to physically connect them together. The 90:10 fibre coupler gives us an estimate of the pump power that we are sending to the chip. The characterisation for both of these components is shown in Fig.~\ref{fig:Grating_Coupler_Spectrum}, and tells us that our loss per grating coupler is 3.75~dB (which is what we expect from this standardized component), and our 90:10 fibre coupler is operating as an 89:11 coupler.

\begin{figure}[h!]
	\centering
	\includegraphics[width=0.8\columnwidth]{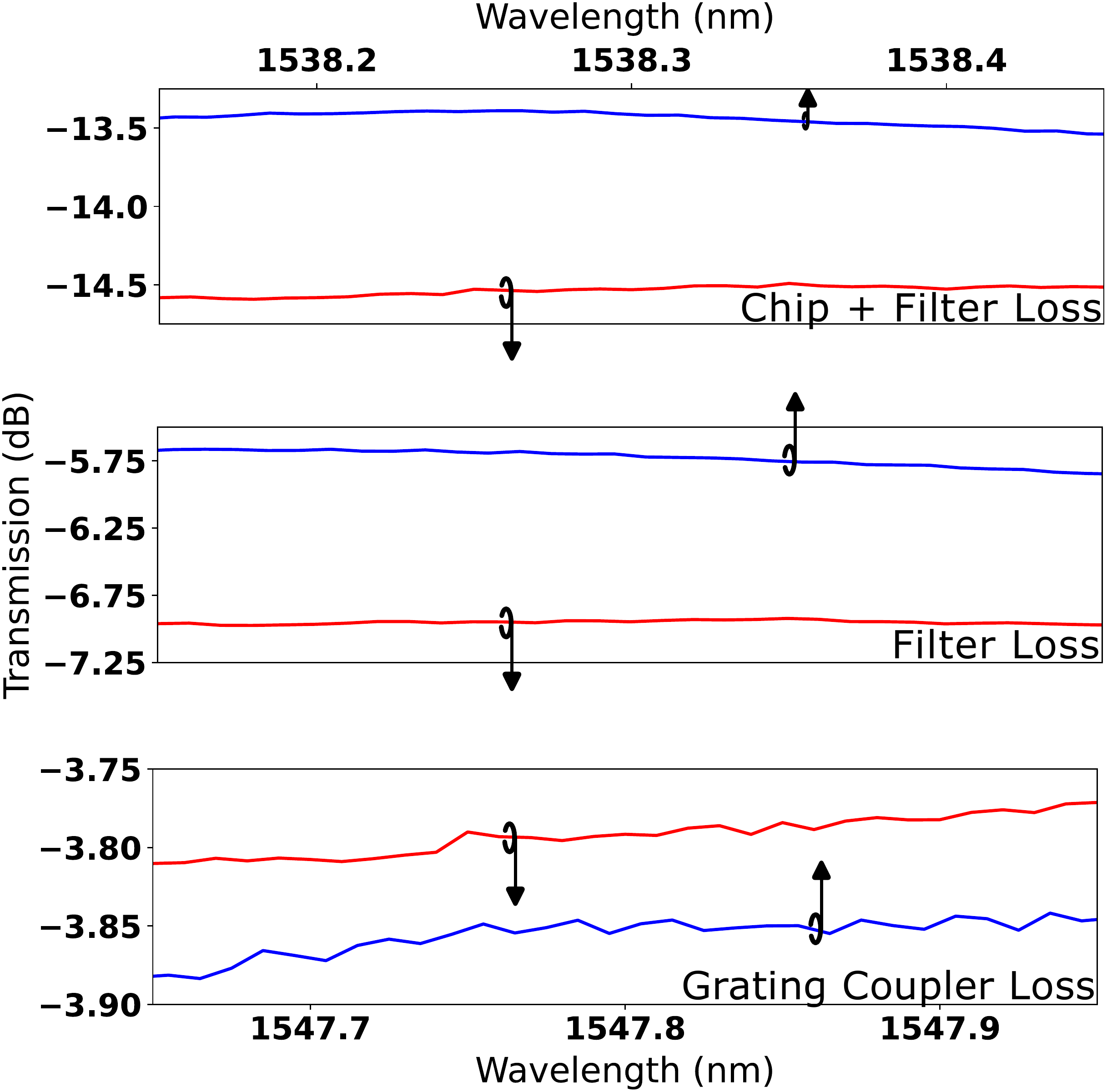}
	\caption{Transmission spectra of the signal (blue) and idler (red) channels for different parts of the experiment. Total transmission of the chip and the filter network (top), the filter network (middle), and the calculated transmission of a single grating coupler (bottom).}
	\label{fig:PM_GC_Loss}
\end{figure}

After this preliminary characterization, we move on to the source structure itself. We need to measure the experimental setup in two configurations, once through the chip and the entire filter network, and once with just the filter network. The results of these measurements are presented in Fig.~\ref{fig:PM_GC_Loss}. This lets us reconfirm our measurements of the grating coupler loss that we got earlier (Fig.~\ref{fig:Grating_Coupler_Spectrum}), which we can see agree well.

\begin{figure}[h!]
	\centering
	\includegraphics[width=\columnwidth]{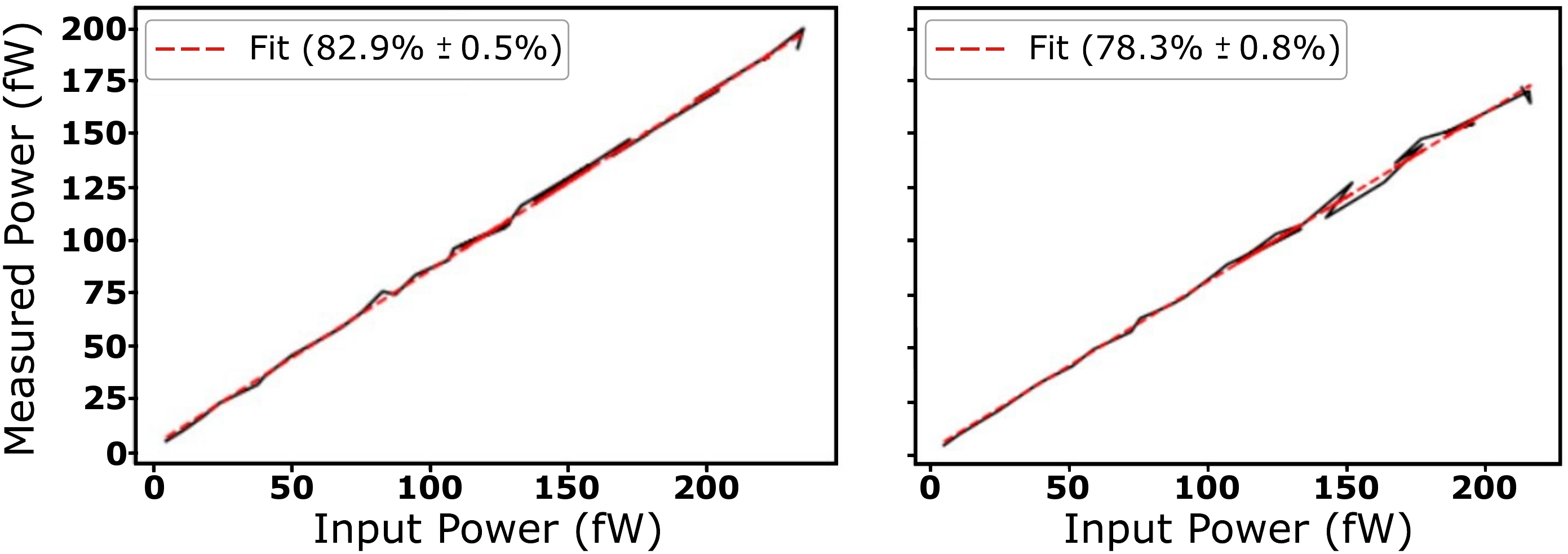}
	\caption{Measurements of system detection efficiencies for the idler detector (left) and signal detector (right). Courtesy of Stefano Paesani and Laurent Kling.}
	\label{fig:Detection_Efficiencies}
\end{figure}

\begin{figure}[h!]
	\centering
	\includegraphics[width=0.8\columnwidth]{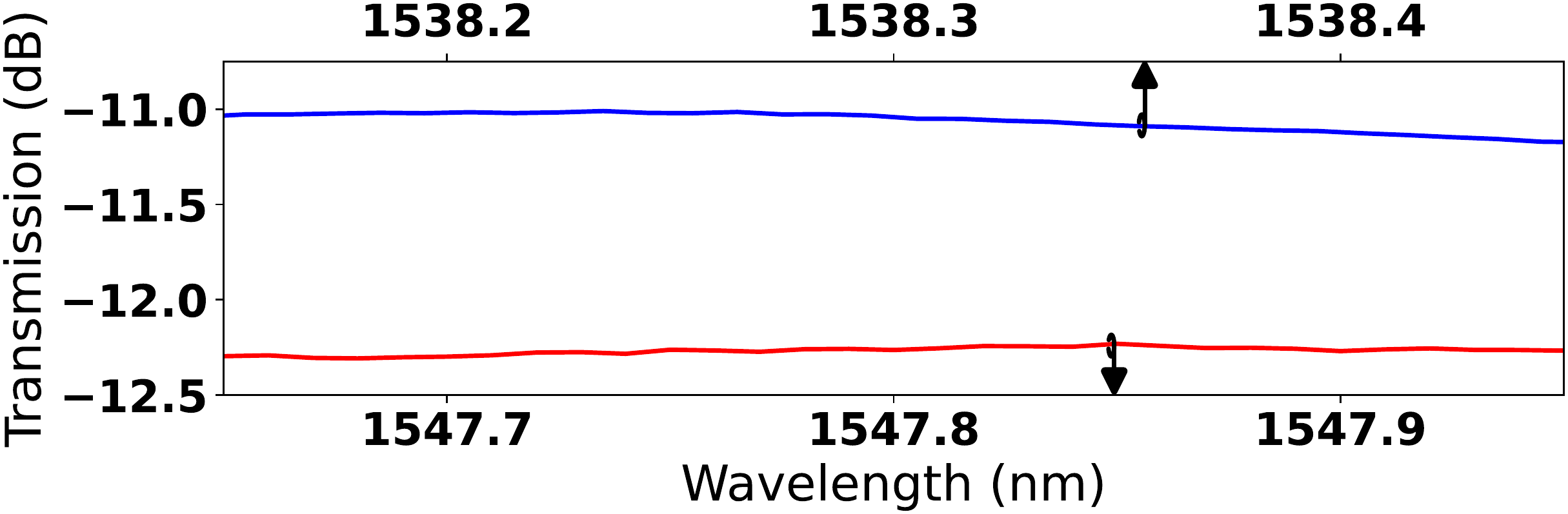}
	\caption{Calculating the loss that the signal (blue) and idler (red) channels experience, from grating couplers to detector efficiencies.}
	\label{fig:PM_GC_Loss_Off_Chip}
\end{figure}

This leaves us with two remaining loss measurements, from the filters to the detector system, and the detection efficiency of the detectors themselves. Using a commercial optical power meter, we measured the transmission loss between the filters, and the detector system to be -0.42~$\pm$~0.02~dB and -0.71~$\pm$~0.02~dB for the signal and idler channels, respectively. As for the detector characterization, this work was done earlier by Laurent Kling and Stefano Paesani who we acknowledged in the main text. The outcomes of these measurements are shown in Fig.~\ref{fig:Detection_Efficiencies}. Giving us a final loss per channel shown in Fig.~\ref{fig:PM_GC_Loss_Off_Chip}

Subtracting these losses from the calculated heralding efficiencies, we see source heralding efficiencies of $\eta_s$ = 92.1~$\pm$~3.2~\%, and $\eta_i$ = 94.0~$\pm$~2.9~\% as in the main text.

\clearpage
\section{Assumptions in Approximating the JSA as the $\sqrt{JSI}$}

\begin{figure}[h!]
	\centering
	\includegraphics[width=\columnwidth]{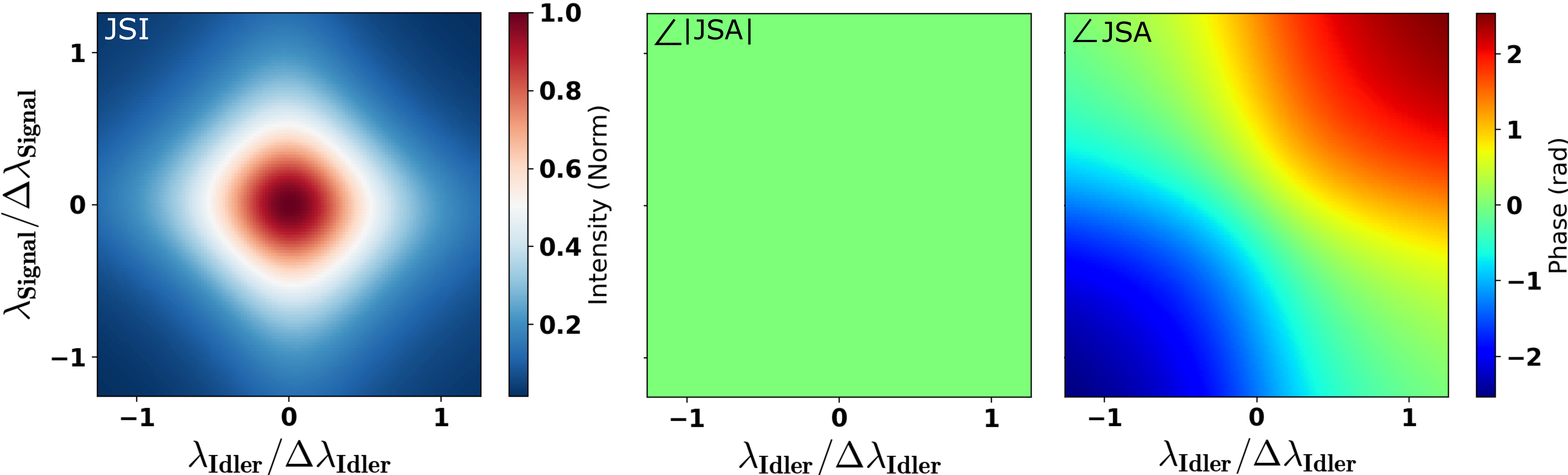}
	\caption{Left) Plotting the simulated JSI of the photonic molecule. Middle) Plotting the joint spectral phase (JSP) of the simulated $|$JSA$|$. Right) Plotting the JSP of the simulated JSA.}
	\label{fig:JSI_JSA_JSP}
\end{figure}

Our assumption that we made in the main text comes from the fact that we typically perform intensity measurements of the JSA using SET, rather than phase-sensitive ones. In our case, phase-sensitive measurements are unnecessary because, if we simulate the JSA and the $|$JSA$|$ the difference in purity is only 0.01\% which is within the error of our experimental data. The difference between these two simulations is simple, in one case the phase of the JSA ($\angle{JSA}$) is non-zero, and in the case of the $|$JSA$|$ it is neglected and assumed to be zero (Fig.~\ref{fig:JSI_JSA_JSP}).

\clearpage

\section{Error Analysis of the Experimental JSI}

Analyzing the experimental JSI properly is important to be able to say with certainty that the estimated purity is a good representation of the data. In our case, we did this by making use of the high-resolution of our equipment (0.16~pm, and 1~pm for the idler and signal axis, respectively). By averaging across several pixels in the JSI, we can reduce the error due to experimental noise. Mostly, experimental noise comes from the accuracy of our optical spectrum analyzer (OSA), and the precision to which we can set the wavelength of our very narrow linewidth seed laser. We can see that by the time we are at a resolution of $\sim 4x4~pm$, the purity value of the JSI is plateauing (Fig.~S14), and the error is minimized. This means that we can say with certainty that the value we quote for the purity of our photonic molecule, is representative of its real-world value. 

\begin{figure}[h!]
	\centering
	\includegraphics[width=\columnwidth]{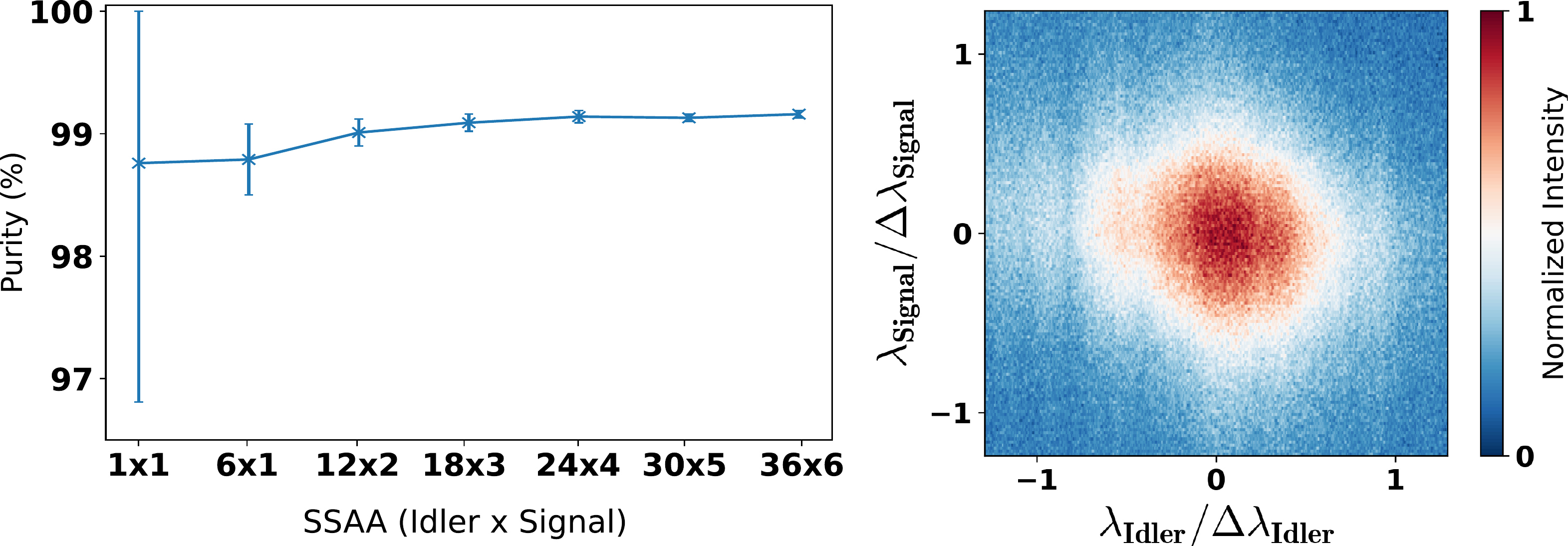}
	\caption{Left) Plotting the dependence of purity with the amount of supersampling. The degree of SSAA indicates the number of data points averaged over. Right) The experimentally sampled JSI, with an idler resolution of 0.16~pm and signal resolution of 1~pm.}
	\label{fig:Error_minimisation}
\end{figure}

As for how we calculate the error on our JSI, we can take a look at the noise distribution of our high-resolution JSI. We can isolate the uncertainty on each data point of the JSI by measuring the variance of each pixel when compared to its neighbor. This works because the noise varies much quicker than the JSI, on the order of individual data points (looking at the speckle on the JSI in Fig.~S14) rather than on the order of thousands of pixels. This gives us an error of $\pm$4\% on each data point. We can work out how sensitive the JSI is to noise by using Monte Carlo simulations and our estimated error on each pixel. This gives us the error that we use in the main text, and in Fig.~S14.

\end{document}